\documentclass[aps, prl, reprint,superscriptaddress, longbibliography]{revtex4-2}

\usepackage{graphicx}
\usepackage{hyperref}
\hypersetup{
    bookmarks=false,         
    unicode=false,          
    pdftoolbar=false,        
    pdfmenubar=true,        
    pdffitwindow=false,     
    pdfstartview={FitH},    
    pdftitle={},    
    pdfauthor={},     
    pdfsubject={},   
    pdfcreator={},   
    pdfproducer={}, 
    pdfkeywords={quantum rotors} {topological physics}, 
    pdfnewwindow=true,      
    colorlinks=true,       
    linkcolor=black,          
    citecolor=magenta,        
    filecolor=magenta,      
    urlcolor=magenta
}

\usepackage[normalem]{ulem}
\PassOptionsToPackage{usenames,dvipsnames}{xcolor}
\usepackage{color}
\usepackage{bm}
\usepackage{amsmath, amsthm, amssymb}
\usepackage{mathtools, xfrac}
\usepackage{mathdots}
\usepackage{algorithm}
\usepackage{algpseudocode}
\usepackage{bm}

\definecolor{commentcolor}{RGB}{0,128,0}
\definecolor{keywordcolor}{RGB}{0,0,255}
\definecolor{stringcolor}{RGB}{163,21,21}

\newcommand{\bk}{\bm{k}}

\newcommand{\with}{\quad \text{with}\quad}

\renewcommand{\and}{\quad \text{and}\quad}
\newcommand{\ie}{i.\,e. }

\renewcommand{\vec}{\mathbf}
\renewcommand{\d}{\mathrm{d}}

\newcommand{\im}{\mathrm{i}\mkern1mu}

\newcommand{\sub}[1]{{\raisebox{-0.5ex}{$\scriptscriptstyle #1$}}}
\begin{document}
\title{Anomalous multi-gap topological phases in periodically driven quantum rotors}
\author{Volker Karle}
\email{volker.karle@ista.ac.at}
\author{Mikhail Lemeshko}
\email{mikhail.lemeshko@ista.ac.at}
\affiliation{Institute of Science and Technology Austria, Am Campus 1, 3400 Klosterneuburg, Austria}
\author{Adrien Bouhon}
\author{Robert-Jan Slager}
\email{rjs269@cam.ac.uk}
\author{F.~Nur Ünal}
\email{fnu20@cam.ac.uk}
\affiliation{TCM Group, Cavendish Laboratory, Department of Physics, JJ Thomson Avenue, Cambridge CB3 0HE, United Kingdom\looseness=-1}

\begin{abstract}
We demonstrate that periodically driven quantum rotors provide a promising and broadly applicable platform to implement multi-gap topological phases, where groups of bands can acquire topological invariants due to non-Abelian braiding of band degeneracies.
By adiabatically varying the periodic kicks to the rotor we find nodal-line braiding, which causes sign flips of topological charges of band nodes and can prevent them from annihilating, indicated by non-zero values of the 
patch Euler class. In particular, we report on the emergence of an anomalous Dirac string phase arising in the strongly driven regime, a truly out-of-equilibrium phase of the quantum rotor. This phase emanates from braiding processes involving all (quasienergy) gaps and manifests itself with edge states at zero angular momentum. Our results reveal direct applications in state-of-the-art experiments of quantum rotors, such as linear molecules driven by periodic far-off-resonant laser pulses or artificial quantum rotors in optical lattices, whose extensive versatility offers precise modification and observation of novel non-Abelian topological properties.
\end{abstract}
\maketitle

{\it Introduction.}--- Over the past decade, research in topological physics has transitioned from primarily investigating topological insulators and semi-metals~\cite{Rmp1, Rmp2, Clas1, Clas2, Clas3, Clas4, Clas5, Slager2019, Shiozaki14, Weylrmp} to a broader exploration of platforms~\cite{Kitagawa10_PRB,Rudner13_PRX,RoyHarper17_PRB, Unal2019}. 
Recently, multi-gap phases~\cite{BJY_nielsen,bouhon2019nonabelian, bouhon2020geometric, bouhon2018wilson, naHopf2024,Ahn2019, 3+1paper, rHopf} have emerged as a new class of topological matter that generically goes beyond single-band characterizations. In such systems groups of bands (band subspaces) can acquire multi-gap invariants~\cite{BJY_nielsen,bouhon2019nonabelian, bouhon2020geometric, bouhon2018wilson, naHopf2024,Ahn2019, 3+1paper, rHopf}. 
The most notable example is Euler class arising in systems with minimum $N\geq 3$ bands featuring a real Hamiltonian due to the presence of the $\cal{PT}$ (combination of parity and time reversal) or $C_2\cal{T}$ (two-fold rotations and time reversal) symmetry, which bestows non-Abelian frame charges to band singularities~\cite{bouhon2019nonabelian, Wu1273, BJY_nielsen}. These charges can be changed by braiding nodes residing between adjacent pairs of bands in momentum space. The braiding can thus result in nodes with the same charges in a two-band subspace (`gap') that cannot annihilate each other, which is a physical quantity captured by the Euler invariant. 
While still a nascent field, multi-gap topologies are increasingly finding their way to a multitude of settings, including phononic~\cite{Peng2021,Peng2022Multi}, electronic~\cite{bouhon2019nonabelian,Koneye2021,chen2021manipulation} and magnetic systems~\cite{magnetic}, metamaterials~\cite{Jiang2021, jiang2024,jiang21} as well as out-of-equilibrium contexts~\cite{Unal_quenched_Euler, zhao2022observation,slager2022floquet,LiHu23_NatComm_floq,Slager_ADS_comment,breach2024interferometry} including an anomalous variant necessitating periodic driving~\cite{slager2022floquet}.

On another front, it has recently been shown that periodically kicked quantum rotors can exhibit Floquet topological phases~\cite{PhysRevA.77.031405,PhysRevLett.109.010601,PhysRevA.97.063603,PhysRevE.96.022216}. This platform, in particular, is interesting given that these systems, which are driven far from equilibrium, are markedly different from solid-state systems, yet they offer clean settings with a number of knobs to control for realizing novel condensed matter phenomena. 
A remarkable example as a clear advantage in this regard is the number of (Floquet) bands $N$ which is governed by the number of kicks within a single revival time of the rotor~\cite{izrailev1980quantum, santhanam2022quantum,wimberger2014nonlinear} and can be selected arbitrarily. This flexibility lays the groundwork for the exploration of hitherto uncharted multi-band topological phenomena and particularly those out of equilibrium. 
The time evolution of the quantum rotor, induced through the periodic driving, can harbour topological charges (singularities) that manifest as significant changes in observables, such as the orientation of a rotating molecule~\cite{PRL_charges}. These changes are protected by an inversion symmetry, $\mathcal{P}$. We here show that preserving time-reversal symmetry ($\mathcal{T}$) in kicked quantum 
rotors allows for these topological singularities to have non-Abelian properties for $N\geq 2$~\cite{bouhon2019nonabelian, Wu1273, BJY_nielsen}.

We investigate non-Abelian braiding of topological singularities by periodically kicking quantum rotors. Upon tuning the strength of the laser pulses (kicks), we demonstrate that non-Abelian charges of band nodes can be changed by moving them around each other while the rotor is evolving. In particular, this allows us to realize an anomalous Dirac string phase with no equilibrium counterpart~\cite{slager2022floquet}, in which {\it all} gaps in the Floquet spectrum of the rotor, including the anomalous $\pi$-gap, harbour topological singularities and contribute to the braiding process. Importantly, these multi-gap phases manifest themselves with  zero-angular-momentum edge states  as smoking-gun signatures. 
Our findings open a new avenue for current experimental studies of quantum rotors, spanning a broad range including linear molecules driven by far-off-resonant laser pulses, quantum nanorotors and synthetic quantum rotors~\cite{koch_quantum_2019, FlossPRE15, bitter_control_2017,PhysRevLett.109.010601,nanorotors}. The adaptability of these experiments, where even the number of bands can be controlled, allows for extensive tuning of system parameters, e.g.~by adjusting the laser strength, providing a new platform for non-trivial multi-band topology.

\begin{figure*}[htp]
\includegraphics[width=0.80\textwidth]{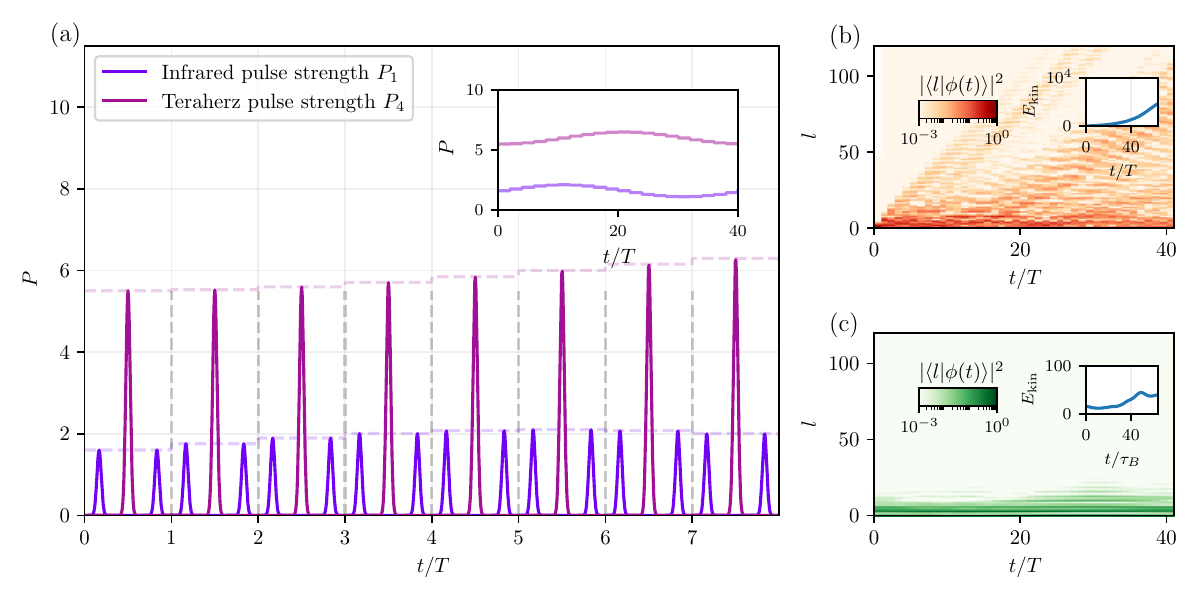}
\caption{Triple-kicked rotor~\eqref{eq:TKR} with three bands ($N=3$) and $P_2=P_3=0$, which reduces the protocol to two infrared pulses and one terahertz pulse within each Floquet period. We vary the pulse strengths stroboscopically on a circular path,
$P_1(\alpha) = 1.6 + \sin(\alpha)/2$, $P_4(\alpha) = 6.0 - \cos(\alpha)/2$,
where $\alpha \in [0,2\pi]$ serves as our synthetic second (momentum) dimension. (a) Pulse sequence as a function of Floquet period. 
The inset shows the variation of pulse strengths over the entire protocol, spanning $N_\gamma=40$ iterations with three pulses per sequence within each of which $\alpha$ is adjusted as $\alpha_n = 2\pi n/N_\mathrm{\gamma}$. In (b) and (c) we show the full time evolution of a driven rotor under this protocol [Eq.~\eqref{eq:StateEvolution}]. (b) Starting with typical thermal state (with $\pi/(\tau_B k_B T)=0.17$), we observe resonant energy absorption ($E_\mathrm{kin} = \langle \mathbf{\hat{L}}^2 \rangle \approx 5\cdot 10^3$), as expected for a generic initial state in this system. Conversely, (c) showcases an edge state (with the same initial energy) that persists throughout the protocol and absorbs little energy ($\langle \mathbf{\hat{L}}^2 \rangle \approx 50$). Significantly, this is an anomalous $\pi-$gap topological edge state, witnessing the Dirac String phase depicted in Fig.~\ref{fig:fig2}(d)), and exists even though all bulk Zak phases vanish [see text].
}
\label{fig:fig1}
\end{figure*}

{\it Model.}--- We consider three-dimensional quantum rotors, such as linear molecules, periodically driven by ultrashort off-resonant pulses~\footnote{These pulses instigate Raman transitions, leading to transitions between different angular momentum states. The pulse power and duration, rather than the energy difference (which increases linearly for angular momenta), are significant, as the pulses are off-resonant. This simplifies the construction of an artificial lattice without the need for multiple pulses with different energies.}. The Hamiltonian of the free rotor can be expressed in terms of the squared angular momentum operator, $H_0= \pi \hat{\mathbf{L}}^2 / \tau_B$, with the rotational period $\tau_B$. We set $\hbar=1$ throughout the text. For pulses significantly shorter than the rotor's rotational period~\footnote{In~\cite{PhysRevA.109.023101}, it has been shown explicitly under which circumstances the sudden-pulse approximation is applicable.}, the effective potential for one alignment pulse can be represented as~\cite{cai2001recurring, dion_laser-induced_1999,PhysRevA.109.023101} $\hat{V}(P_1,P_2)=P_1 \cos(\hat{\theta}) + P_2\cos^2(\hat{\theta})$, with pulse strengths $P_1, P_2$, and the angle $\hat{\theta}$ between the axis of the rotor and the linear polarization of the laser. As this potential does not alter the magnetic quantum number $m$, we focus on the case $m=0$~\footnote{An extension of the one-dimensional lattice in $l$ to a two-dimensional case incorporating $m \in \{-l,\dots,l\}$ is straightforward. However, this extension shall not be addressed here, as it would introduce additional complexity to the phenomenon of interest discussed in this work and left for exploring in future research.}. 
In this picture, angular momentum states $l\geq0$ serve as lattice sites in a semi-infinite lattice, tunneling among which is controlled by the periodic laser pulses~\cite{PRL_charges}. 

\begin{figure*}[htbp!]
\centering
\hspace{-0.97cm}
    \includegraphics[width=1.05\textwidth]{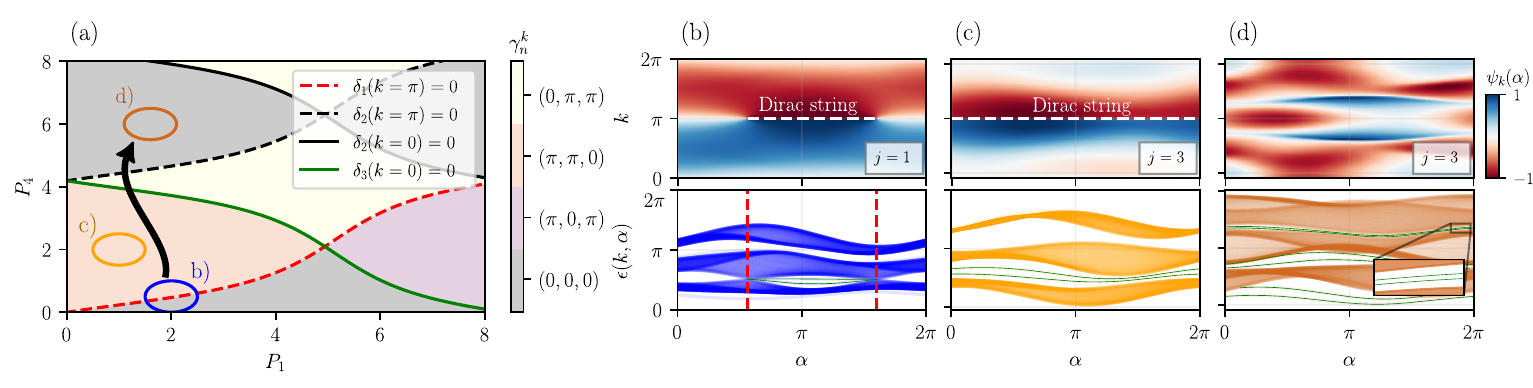}
    \caption{
    (a) Phase diagram for the triple-kicked rotor with $P_2=P_3=0$, with Zak phases of each band $\gamma_n^k$ along $k-$direction. We identify four different configurations, separated by nodal lines at $k\in \{0,\pi\}$ where the gap $\delta_n$ in Eq.~\eqref{eq:gap_function} vanishes. Additional nodal lines at arbitrary $(k,-k)$ are not shown for clarity (see Ref.~\cite{sup} for all phase crossings). Colored circles represent the cyclic modulation in $\vec{P}(\alpha)$ [c.f.~Fig.~\ref{fig:fig1}], which results in the formation of a DS if it crosses over a nodal line as in (b). (b-d) the $j$th component of the $n=2$ eigenstate (upper panels) and the corresponding quasienergy spectrum (lower panels) along the path marked by an arrow in (a) traversing through several topological phase transitions. (c) Two nodals points annihilate across the BZ, leaving their DS behind, which changes the Zak phase along the perpendicular direction to $\gamma_1^k\!=\!\gamma_2^k\!=\!\pi$ and induces edge states in this gap with open boundary conditions.    
    (d) After crossing two more nodal lines in the two remaining gaps, the Zak phase of each band is set back to zero thanks to the nodal line in the anomalous gap ($\delta_3$) enabled by periodic kicking. This is an anomalous Dirac string phase with edge modes in each gap, despite the trivial Zak phases which are the band invariants at equilibrium and cannot capture this topological regime on their own. 
}
    \label{fig:fig2}
\end{figure*}

We employ Floquet theory to analyze the one-period time-translation operator~\cite{drese1999floquet,PhysRevA.109.023101}, 
\begin{equation}
U_\mathrm{KR}(P_1,P_2)=e^{-i\pi\hat{\mathbf{L}}^2 \tau /(2\tau_B)}e^{i \hat{V}(P_1,P_2)}e^{-i\pi \hat{\mathbf{L}}^2 \tau /(2\tau_B)},
\end{equation}
where the period between the pulses is $\tau$. We here consider the explicitly time-reversal invariant Floquet gauge with the kick applied in the center~\footnote{The Floquet operator has a gauge freedom in when the kick occurs. In the gauge with the single kick in the center, the system is time-reversal invariant with $\mathcal{T}=1$. 
It is straightforward to show that for any time-reversal operator in this gauge $\tilde{U}_1, \dots, \tilde{U}_n$, the product $\tilde{U}_1 \dots \tilde{U}_{n-1}\tilde{U}_n \tilde{U}_{n-1}\dots \tilde{U}_1$ also has time-reversal symmetry.}. 
Furthermore, we focus on the quantum resonance case with $\tau=2\tau_B/N$ for $N\geq 3$, where the system exhibits $N$ quasi-energy bands~\footnote{More strictly, there are a maximum of $N$ energy bands up to accidental symmetries for odd $N$, and for even $N$, there are $2N$ energy bands. The number of bands originates from the periodicity in the lattice of the free rotation
$\langle l'| e^{-i \tau \hat{L}^2/\tau_B}|l \rangle = e^{-i \pi l(l+1)/ N } \delta_{ll'}$ as demonstrated in Ref.~\cite{PRL_charges}.}. 

The laser pulses introduce hopping in the synthetic lattice formed by the angular momentum states, which results in an effective tight-binding model~\cite{PRL_charges} and allows us to define the quasi-momentum $k\in [0,2\pi]$~\footnote{See supplementary material for a more comprehensive derivation of the Fourier transform~\cite{sup}.}. 
Depending on the kicking strength, there is tunneling between not only two nearest-neighbor sites but also next-to-nearest and further neighbors, which is indeed crucial for the existence of non-trivial nodal lines in the system~\cite{Unal_quenched_Euler}.

For the kicked rotor, the stroboscopic Hamiltonian, $H_\mathrm{KR}=i\log(U_\mathrm{KR})/\tau$, with Floquet states $\psi_n(k)$ and quasi-energies $\epsilon_n(k)\tau\in [-\pi,\pi]$ reads
\begin{equation}
H_\mathrm{KR}(k) \psi_n(k) = \epsilon_n(k)\psi_n(k),
\end{equation}
where $n \in \{1 \dots N\}$ labels the band index. We, hereafter, absorb the Floquet period in the definition of $H_\mathrm{KR}$ and quasi-energy for simplicity. 
Resolved as a phase, the $2\pi$ periodicity of the Floquet spectrum allows for novel topological phenomena, including gap closings in the so-called $\pi-$gap between bands $n=1$ and $n=N$. This gives rise to anomalous Floquet topologies with no equilibrium counterpart which have been attracting great attention as they offer unique properties~\cite{Kitagawa10_PRB,Rudner13_PRX,Wintersperger20_NatPhys,slager2022floquet,Titum16_PRX_AFAI,Nathan19_PRB_AFI}.

We emphasise that $H_\mathrm{KR}$ obeys an inversion symmetry $\mathcal{P}$ and time-reversal symmetry $\mathcal{T}$ with
\begin{equation}
\mathcal{P}H_\mathrm{KR}(k) = H_\mathrm{KR}(-k)\mathcal{P}, \quad \mathcal{T}H_\mathrm{KR}(k) = H^*_\mathrm{KR}(-k)\mathcal{T},
\label{eq:Symmetry_Hkr}
\end{equation}
which allows us to adopt a real, Euler, definition in the subsequent. As was shown in Ref.~\cite{PRL_charges}, the single kicked rotor exhibits only linear nodal lines. Since the double-kicked rotor in general breaks time-reversal symmetry, we consider a triple kick of the form
\begin{equation}
U_{\mathrm{TKR}}(\mathbf{P}) = {U}_{\mathrm{KR}}(P_1,P_2){U}_{\mathrm{KR}}(P_3,P_4){U}_{\mathrm{KR}}(P_1,P_2),
\label{eq:TKR}
\end{equation}
as the simplest model to demonstrate the braiding of nodal lines.
Here, the Floquet period becomes $T=3\tau=3\tau_B/(2N)$ as the relevant stroboscopic time-scale in the remainder of the discussion. 
The triple-kicked rotor~\eqref{eq:TKR} has four parameters $\boldsymbol{P}=(P_1, \dots , P_4)$ for the strength of the kicks that we control to induce band touching points between adjacent bands. 

{\it Quantum Rotor and anomalous Dirac string phase.}--- The triple-kicked rotor Hamiltonian ($H_{\mathrm{TKR}}$) can be cast into a real form due to the $\cal{PT}$ symmetry established in Eqs.~\eqref{eq:Symmetry_Hkr}--\eqref{eq:TKR}. Focusing on the $N=3-$band resonance case, the space of the flattened Hamiltonians is given by $O(3)/O(1)^3=O(3)/\mathbb{Z}_2^3 = SO(3)/D_2$ where the eigenstates form a real and orthonormal frame, a `dreibein', upon fixing its orientation~\cite{bouhon2019nonabelian}. While the two eigenstates involved in a band touching point are not uniquely defined at the node, they pick up a $\pi$-Berry phase upon circling around it, corresponding to also the frame rotating by $\pi$~around an axis fixed by the two-band subspace \cite{breach2024interferometry}. These nodes can be characterized by a global frame charge taking value in the first homotopy group $\pi_1(\mathrm{SO(3)/D_2}) = \mathbb{Q}$, that is the non-Abelian quaternion group $\mathbb{Q}=\{+1,\pm \mathbf{i}, \pm \mathbf{j}, \pm \mathbf{k},-1 \}$~\cite{Wu1273}. Most importantly, the sign of the node's frame charge can be reversed upon braiding with a node hosted in either adjacent gaps (i.e.~the gap directly above or below), hence directly reflecting the anticommuting algebra of the quaternion elements~\cite{bouhon2019nonabelian,bouhon2020geometric,slager2022floquet}.


The non-Abelian braiding of band singularities naturally necessitates a two-dimensional parameter space hosting the braided loops. In kicked quantum rotors, while quasi-momentum $k$ corresponds to the first dimension, we establish a second dimension by modulating the kicking strengths over time in a periodic way as a function of $\alpha=\alpha(t)\in [0,2\pi]$ as illustrated in Fig.~\ref{fig:fig1}a for a $N=3-$band model, $H_{\mathrm{TKR}}(k,\alpha)$. 


Under this protocol, an initial state $| \phi_0 \rangle$ evolves as 
\begin{equation}
   |\phi_{N_\gamma} \rangle = U_\mathrm{TKR}(\mathbf{P}(\alpha_{N_\gamma}))\cdots U_\mathrm{TKR}(\mathbf{P}(\alpha_1))|\phi_0\rangle ,
   \label{eq:StateEvolution}
\end{equation}
where $N_\gamma$ denotes the number of stroboscopic iterations completing a cycle for $\alpha_{N_\gamma}=2\pi$. 
By tuning the kicking strengths, we observe both extended bulk and localized edge states as shown in Fig.~\ref{fig:fig1}(b) and (c) respectively. As expected for a quantum rotor kicked at quantum resonance, a generic thermal state $\phi_0(l) \propto \exp(- \pi l(l+1)/(\tau_B k_B \mathrm{T}))$ at temperature $\mathrm{T}$ absorbs substantial energy from the driving over time and spreads in momentum space. In contrast, an edge state with identical initial energy remains localized under the same driving protocol. 
By adiabatically connecting to the static regime~\cite{Unal2019,slager2022floquet,Martinez23_wp}, we identify this edge state as anomalous, residing in the $\pi-$gap [see Fig.~\ref{fig:fig2}(d) for the corresponding spectrum]. Crucially, it exhibits topological protection as it can only be eliminated through gap closure and braiding. We will now elucidate this relationship and topological characterization in greater detail.


\begin{figure*}
  \begin{minipage}[b]{0.60\textwidth}
    \includegraphics[width=1\linewidth]{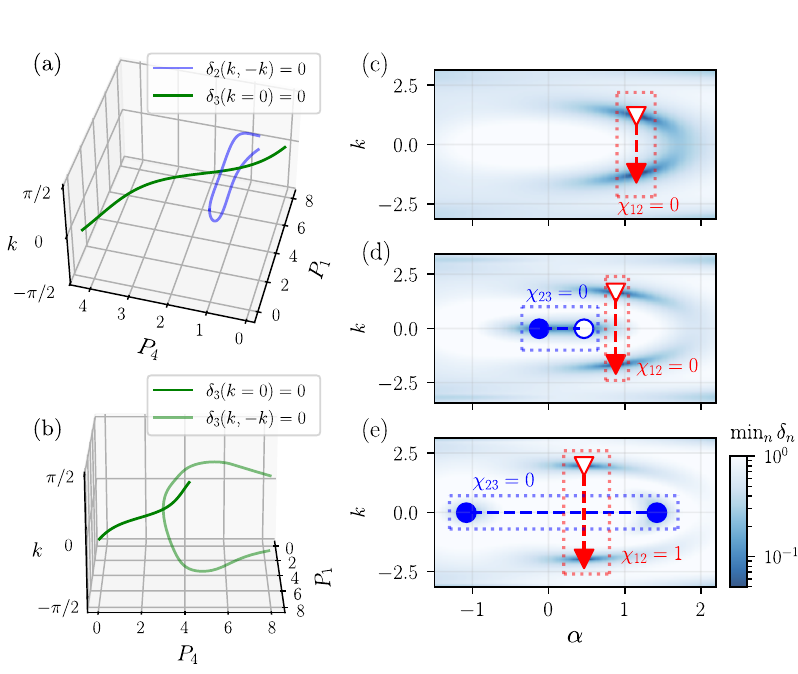}
  \end{minipage}
  \hfill
  \begin{minipage}[b]{0.39\textwidth}
    \caption{(a) Non-Abelian braiding in $(k,\alpha)$ space with the same parameters as in Fig.~\ref{fig:fig2}. Examples of nodal lines, showing how the nodal line in gap 2 (blue line) winds around the nodal line of gap 3 (green) during braiding. (b) Formation of a $(k,-k)$ nodal line around a $k=0$ nodal line. Other nodal lines are not shown for better visibility~\cite{sup}. (c-e) The triple-kicked quantum rotor can realize braiding processes and non-trivial patch Euler class for various parametrizations, here shown over a different path $P(\alpha, \beta)=f(\cos(\alpha),\sin(\alpha),\beta)$ with some smooth function $f(x,y,\beta)$ interpolating between different positions of band nodes (see text). The minimal gap size in the background reveals (c) a pair of nodes with opposite charges (depicted with empty/filled triangles, $\chi_{12}=0$) in gap 1 for $\beta=0.15$, and (d) one more pair with opposite charges in the adjacent gap (circles) for $\beta=0.21$. (e) Moving a band node across the DS (dashed line) in the adjacent gap flips its sign, here $\beta=0.3$. The nodes in gap 1 (triangles) now cannot annihilate within the red dotted area as one has to go through the DS in gap 2, changing their sign, which is captured by the non-trivial Euler class $\chi_{12}=1$ within this patch.    
    }
    \label{fig:fig3}
  \end{minipage}
\end{figure*}

An equivalent way of tracking the non-Abelian braiding in the multi-gap Euler topology involves Dirac strings (DSs)~\cite{Peng2021,Jiang2021,slager2022floquet}. In general, the gauge of the eigenstates cannot be chosen smoothly across the whole momentum space $\bk=(k,\alpha)$ in the presence of spectral degeneracies whose $\pi$--Berry phase imposes a line (DS) along which the eigenstates change their sign, as illustrated in Fig.~\ref{fig:fig2}(b). While DSs are gauge objects, i.e.~their exact position in $\bk$ cannot be fixed, crossing a DS induces a $\pi$ Zak phase $\gamma_n = \oint \langle \psi_n(\bk)|\nabla_k \psi_n(\bk)\rangle\!\cdot\!\mathrm{d} \bk$, which is calculated along non-contractible paths and indeed a gauge-invariant observable quantity~\cite{Zak89_PRL}. Similarly, also the sign of a frame charge changes upon crossing a DS lying in an adjacent gap [see Fig.~\ref{fig:fig3}(d-e)], successfully encoding the braiding of band nodes~\cite{slager2022floquet,breach2024interferometry}. We note that time-reversal symmetry and inversion in our system enforces that nodal points come either in pairs at $(-k,k)$ connected by DSs, or occur at high-symmetry points $k=0$ or $k=\pi$~\cite{PRL_charges}. Namely, the nodal points carry topological charges and can only be created or annihilated in pairs.

By tuning pulse strengths in the triple-kicked rotor we create band nodes and move them around in the $\bk$-space before annihilating them across the Brillouin zone (BZ), leaving their DS behind as shown in Fig.~\ref{fig:fig2}(b-c). This manifests as a $\pi$
jump in the Zak phase along the orthogonal direction, $\gamma_n^k$, in bands $n\!=\!1$ and $n\!=\!\!2$, and we observe the corresponding edge states in gap 1~\footnote{Edge states are localized at two distinct boundaries: $l=0$ and $l=l_\mathrm{max}$, where $l_\mathrm{max}$ is system dependent. Notably, these edge states are not degenerate due to the system's imperfect translational invariance and particularly minor inhomogeneities at the boundaries as detailed in~\cite{sup}.}. Fig.~\ref{fig:fig2}(a) demonstrates the phase diagram with the Zak phases of each band marked. Zak phases of bands change upon crossing each nodal line, of which we keep track by defining a gap function (modulo $2\pi$~\footnote{Here we define for $x,y \in \mathbb{R}$ the $2\pi$-remainder norm $$|x - y|\sub{2\pi} = \min_{n \in \mathbb{Z}}|x-y+ 2\pi n| $$ which gives the minimal distance on the circle between radiants $x,y$.}),
\begin{equation}
\delta_\mathrm{n}(\bk,\vec{P})=|\epsilon_n(\bk,\vec{P}) - \epsilon_{n+1}(\bk,\vec{P}) |_\sub{2\pi} ,
\label{eq:gap_function}
\end{equation}
vanishing at nodal points. We identify four distinct Zak-phase configurations. 

A crucial point underpinned in Fig.~\ref{fig:fig2}(a) is that Zak phases on their own are insufficient to capture the out-of-equilibrium topological phenomena~\cite{slager2022floquet,Slager_ADS_comment}, as they are topological invariants associated with bands only. This stems from the fact that the periodically-kicked quantum rotor can harbor a nodal line also in the anomalous $\pi$--gap ($\delta_3$). Starting from the trivial phase (gray area at the bottom) and following a path traversing various phases (arrow), we arrive at a phase again with vanishing Zak phases upon crossing nodal lines in each gap. However, these two phases (gray-shaded areas) are indeed topologically distinct as one cannot go from one to the other without gap closures~\cite{Slager_ADS_comment}. 
This phase is enabled by the out-of-equilibrium nature of the system and called an anomalous Dirac string (ADS) phase with nodal lines in all gaps in the quasi-energy spectrum~\cite{slager2022floquet}, which can be identified by keeping track of the DSs in a gap-resolved way~\cite{Unal2019}. Indeed, the ADS phase reveals itself in the kicked quantum rotor with edge states in each gap, including the anomalous $\pi$-gap, in Fig.~\ref{fig:fig2}(d). 

{\it Non-trivial patch Euler class}--- 
The periodically kicked rotor offers a rich phase space with abundant possibilities for different nodal-line structures. While we focus on the $k=0$ and $k=\pi$ nodal lines that change Zak phases along the $k$-direction in Fig.~\ref{fig:fig2} to highlight the main phases and in particular the anomalous nature of the system,  $\mathcal{PT}$ symmetry allows also for nodal points at $(k,-k)$ pairs resulting in Zak-phase changes along the $\alpha$-direction as shown in Fig.~\ref{fig:fig3} (see Ref.~\cite{sup} for the full nodal line structure). 
The interplay of nodal lines from different band gaps enables non-Abelian braiding of band singularities, flipping their frame charges when they cross a nodal line in an adjacent gap. 

The obstruction against annihilation of a pair of band nodes between bands $n$ and $n+1$ within a patch $\mathrm{D}\in\bk$ can be captured by the pacth Euler class
\begin{equation}
   \chi_{n,n+1}^\mathrm{D} = \frac{1}{2\pi} \left ( \oint_\mathrm{D} \mathrm{Eu}(\bk) \,\mathrm{d}k \wedge \mathrm{d}\alpha - \oint_{\partial \mathrm{D}}\mathcal{A}\cdot\mathrm{d}\bk \right ) \in \mathbb{Z}.
\end{equation}
Here, the Euler two-form is defined as $\mathrm{Eu}(\bk)=\langle \partial_k \psi_n|\partial_\alpha \psi_{n+1}\rangle - \langle \partial_\alpha \psi_n|\partial_k \psi_{n+1}\rangle$ with corresponding one-form $\mathcal{A}(k,\alpha)=\langle \psi_n |\nabla \psi_{n+1}\rangle$. Nodes after being created in pairs from vacuum [Fig.~\ref{fig:fig3}(c-d)] carry opposite charges, hence, vanishing patch Euler class. However, frame charges associated with nodal lines of adjacent gaps anti-commute and can be changed upon crossing their DSs, here demonstrated by tuning the kicking parameters~\footnote{We choose a parametrization $\alpha(t)$ with a smooth function $f(x,y,\beta)=(1+5\beta(x+1),0.4+2\beta(x+1),0.7+\frac{3\beta}{4}(y+1), 0.7+\frac{3\beta}{2}(y+1))$, noting that this phenomenon is observable for a wide range of parameters and is not a fine-tuned example.}. 
In the resulting configuration [Fig.~\ref{fig:fig3}(e)], the nodes in gap 1 (red triangles) cannot annihilate within the marked patch as the nodal lines have twisted around each other, revealed by a non-trivial patch Euler class.

{\it Conclusion}---
We have demonstrated that periodically kicked quantum rotors provide a versatile platform for realizing multi-gap topological phases. 
By controlling kicking strengths, one can induce nodal lines in each gap, including the $\pi$-gap, culminating in  an anomalous Dirac string phase~\cite{slager2022floquet} in the strongly driven regime. This out-of-equilibrium phase is revealed by zero angular momentum edge states in all gaps. In addition, we have shown experimentally feasible protocols to braid such nodal lines residing between different gaps, which enables non-Abelian braiding of band singularities and, accordingly, a finite patch Euler class.

Our work extends beyond static topological matter, offering insights into novel driven phenomena in the clean, highly controllable setting of quantum rotors. The flexibility in selecting the number of Floquet bands and precise control over system parameters make quantum rotors an ideal test-bed for exploring multi-band topological phenomena. These findings open avenues for future research, including the exploration of systems with $N>3$ bands~\cite{bouhon2020geometric, bouhon2022multigaptopologicalconversioneuler}, the impact of disorder and interactions on topological properties, and the potential for realizing exotic quantum states through engineered non-Abelian braiding processes. Moreover, the direct applicability to state-of-the-art experiments with driven linear molecules and artificial quantum rotors in optical lattices promises immediate opportunities for experimental validation and further exploration of these intriguing topological phases.

\begin{acknowledgments}
	{\it Acknowledgments---}
We thank G. M. Koutentakis, S. Wimberger, J. G. E. Harris, T. Enss and A. Ghazaryan for fruitful discussions. M.L.~acknowledges support by the European Research Council (ERC) Starting Grant No.~801770 (ANGULON). R.-J.~S.~acknowledges funding from a EPSRC ERC underwrite grant EP/X025829/1, a EPSRC New Investigator Award grant EP/W00187X/1, as well as Trinity College, Cambridge. F.N.\"U.~acknowledges support from the Marie Sk{\l}odowska-Curie programme of the European Commission [Grant No.~893915], Simons Investigator Award [Grant No.~511029] and Trinity College Cambridge.
\end{acknowledgments}

\bibliography{main}

\appendix
\begin{widetext}

\setcounter{equation}{0}
\setcounter{figure}{0}
\setcounter{page}{1}
\makeatletter
\renewcommand{\theequation}{S\arabic{equation}}
\renewcommand{\thefigure}{S\arabic{figure}}
\renewcommand{\thetable}{S\arabic{table}}

\section{Appendix A: Fourier transform}
Here we explain the Fourier transform at quantum resonances of periodically driven quantum rotors~\cite{PRL_charges}. At a quantum resonance $T=\frac{\pi}{B \cdot N}$, the time-translational operator takes the form
    $\hat{U} = e^{- \pi \im \hat{\mathbf{L}}^2/ N}e^{\im \hat{V}(P_1,P_2)}$. For odd $N$, the rotational part is periodic with $N$, while for even $N$ the periodicity is $2N$, which becomes evident when looking in the $l$-basis with $\langle l | e^{- \pi \im \hat{\mathbf{L}}^2/ N} | l' \rangle  = e^{-i\pi l(l+1)/N}\delta_{ll'}$. Crucially, while for 2D rotors the matrix elements are constants, for a 3D rotor one has to take into account that the matrix elements depend on $l,l'$. However, they converge to a constant value, \ie $V_{l,l'} \approx V_{l+1,l'+1}$, which is in particular true for the combination of a ``regular'' multi-cycle laser pulse (the $\cos^2(\hat{\theta})$ term) and a half-cycle laser pulse (the $\cos(\hat \theta)$ term), which can be proven using the Edmonds asymptotic formula for the $3j$-symbols~\cite{flude_edmonds_1998}. The asymptotic expressions take the simple form \ie 
    \begin{align}
      \langle l'm'|\cos(\theta)|lm\rangle &= -\delta_{mm'}C^{l'm}_{lm10}C^{l0}_{l'010} \xrightarrow[]{\text{for $l,l'\gg 0$}}\delta_{mm'}\left(\delta_{l,l'+ 1} + \delta_{l,l'- 1}\right)/2 \label{eq:S1}\\
      \langle l'm'|\cos^2(\theta)|lm\rangle &= \delta_{mm'}\left (\tfrac{2}{3}C^{l'm}_{lm20}C^{l0}_{l'020} + \tfrac{1}{3}\delta_{ll'} \right)  
      \xrightarrow{\text{for $l,l'\gg 0$}}\delta_{mm'} \left(\delta_{l,l'} + (\delta_{l,l'+2} +\delta_{l,l'-2})/2\right) /2
      \label{eq:S2}
    \end{align}
    with the usual Clebsch–Gordan coefficients $C^{LM}_{l'm'lm}=\langle l',l,;m',m | l',l; L,M\rangle$. To prove this asymptotic, one can use the Edmonds asymptotic formula for 3j-symbols~\cite{flude_edmonds_1998}  
    \begin{equation}
      {\begin{pmatrix}l_{1}&l_{2}&l_{3}\\m_{1}&m_{2}&m_{3}\end{pmatrix}}\xrightarrow[]{\text{for $l_{2},l_{3}\gg l_{1}$}}(-1)^{l_{3}+m_{3}}{\frac {d_{m_{1},l_{3}-l_{2}}^{l_{1}}(\theta )}{\sqrt {l_{2}+l_{3}+1}}} \with \cos(\theta) = \frac{m_2-m_3}{l_2+l_3+1},
    \end{equation}
    where $d^l_{m,m^\prime}(\theta)$ is the Wigner function and 3j-symbols are related to the Clebsch-Gordan coefficients~\cite{varshalovich_quantum_1988}
    \begin{equation}
\langle j_{1}\,m_{1}\,j_{2}\,m_{2}|J\,M\rangle =(-1)^{-j_{1}+j_{2}-M}{\sqrt {2J+1}}{\begin{pmatrix}j_{1}&j_{2}&J\\m_{1}&m_{2}&-M\end{pmatrix}}.
    \end{equation}
    For Eq.~\eqref{eq:S1} this leads to 
    \begin{equation}
        C^{l'm}_{lm10}C^{l0}_{l'010} = (-1)^{-m}\sqrt{(2l'+1)(2l+1)}\begin{pmatrix}
        1 & l' & l \\
        0 & m & m 
        \end{pmatrix} 
\begin{pmatrix}
        1 & l & l' \\
        0 & 0 & 0 
        \end{pmatrix} 
        \xrightarrow[]{\text{for $l,l'\gg 0$}}(-1)^{l+l'} d^1_{0,l-l'}(\sfrac{\pi}{2})d^1_{0,l'-l}(\sfrac{\pi}{2})
    \end{equation}
    where we used that $\theta = \pi/2$ for both terms.
    Since the Clebsch-Gordan coefficients allow only $l-l'=\pm 1$ and $d^1_{0,\pm 1}(\sfrac{\pi}{2})d^1_{0,\mp 1}(\sfrac{\pi}{2})=\sfrac{1}{2}$ this leads to the desired relation. For Eq.~\eqref{eq:S2} this derivation proceeds analogously, with the difference that  $d^2_{0,\pm 2}(\sfrac{\pi}{2})d^2_{0,\mp 2}(\sfrac{\pi}{2})=\sfrac{3}{8}$ and $d^2_{0,\pm 2}(\sfrac{\pi}{2})d^2_{0,\mp 2}(\sfrac{\pi}{2})=\sfrac{3}{8}$ and $d^2_{0,0}(\sfrac{\pi}{2})d^2_{0,0}(\sfrac{\pi}{2})=\sfrac{1}{4}$, which in total leads to the desired result. For the projection of angular momentum $m=0$ these coefficients converge fast and the overall behavior of the system is very well described by an approximation where we assume a constant value. In this work we only consider $m=0$, but for other $m$ the approximations work out as well when considering larger $l$.
    The $N$-periodicity allows us to define a Fourier transform which approximates the periodic behavior for $l,l' \gg 0$. Any operator $\hat{A}$ with that periodicity, and the effective Hamiltonian in particular, follows
   \begin{equation}
      \langle l' | \hat{A}|l \rangle \approx \langle l'+N | \hat{A}|l+N \rangle 
      \label{eq:period}
   \end{equation}
for $l,l'\gg 0$. Let us parametrize each $l',l$ by $l=n\cdot N + i, l'=n'\cdot N + j$ with $n,n' \in \mathbb{N}_0,i,j \in \{0,1,\dots,N\}$. Then, 
\begin{equation}
  A_{l'l} = A_{i,j}(n',n)=A_{i,j}(n'-n)=A_{i,j}(\Delta n)  
\end{equation}
\ie $A$ only depends on the off-diagonal index $\Delta n$ because of the periodicity. The Fourier transform from $l$- to $k$-space then becomes (for a dimensionless quasi-momentum variable $k \in [0,2\pi]$)
\begin{equation}
\begin{aligned}
\label{eq:Aeff}
    \underbrace{A_{ij}(k',k)}_{N\text{x}N \text{Matrix}} &= \sum_{n',n} e^{-\im ( n'\cdot k' - n \cdot k)}  A_{ij}(n',n) = \sum_{\Delta n, \bar{n}}e^{-\im (\bar{n}(k'-k)/2 + \Delta n (k'+k)/2)} A_{ij}(\Delta n) 
 = \delta_{kk'}\sum_{\Delta n} e^{-\im \Delta n \cdot k}  A_{ij}(\Delta n),  \\
\text{i.e.} \quad A_{ij}(k) &=\sum_{\Delta n} e^{-\im \Delta n \cdot k} A_{ij}(\Delta n) \quad \text{and correspondingly}\quad A_{ij}(\Delta n) =\frac{1}{2\pi} \int_0^{2\pi} A_{ij}(k)\d k.
\end{aligned}
\end{equation}
where we used $\bar{n}=n'+n, \, \Delta n = n'-n$ and $A_{ij}(n',n)=A_{ij}(\Delta n)$, which shows that operator $\hat{A}$ conserves quasi-momentum $k$. 

Since the Fourier transform is only well defined for an infinite system, for a finite system this implies practically an infrared cutoff in quasi-momentum which is defined by the system size. In our case, we are limited by $l=0$ to $l=l_\mathrm{max}$, which in experiment is determined by the validity of the rigid rotor assumption. For example, for linear molecules, at some finite $l_\mathrm{max}$ centrifugal distortion plays a major role and non-linear terms lead to a break-down of the quantum resonance and dynamical localization~\cite{flos_anderson_2014}. When calculating the Fourier transform of an operator, one needs to choose a unit-cell in the middle of the lattice to avoid edge effects. Practically, we choose $A_{ij}(\Delta n) = A_{N \cdot n_0+i,N\cdot (n_0 + \Delta n) +j}$ with a large enough $n_0$ to ensure convergence. For an infinite and perfectly translationally invariant system, we would sum over all $\Delta n \in \mathbb{Z}$. However, here we need to choose $-n_0 \ll \Delta n \ll n_0$. For small to moderate $P$, only few off-diagonals suffice to achieve high agreement with the ``real-space'' diagonalization. A state transforms with
\begin{equation}
   f_i(k) = \sum_{\Delta n} e^{-\im \Delta n \cdot k} f(i + N(n_0 + \Delta n)), \quad f(i+N(n_0 +\Delta n)) =\frac{1}{2\pi} \int_0^{\infty}f_i(k)e^{+\im \Delta n \cdot k}\d k.
   \label{eq:fourier}
\end{equation}
The Fourier transform of the operators at $l',l \gg 0$ in~\eqref{eq:S1} and~\eqref{eq:S2} is given by 
\begin{equation}
  (\cos(\hat{\theta}))_{ij}(k) = \frac{1}{2}\underbrace{\begin{pmatrix}
0 & 1 & & &  & & \dots& & e^{-ik}\\
1 & 0 & 1 &  & & & & & \\
 & 1 & 0 &  & & & & & \\
\vdots  &   &   & \ddots & & & & & \vdots \\
  &  &  &  & & & & 1& \\
 & &  & & & &1 & 0 &1\\
e^{+ik} && \dots & & & & &1 & 0\\
\end{pmatrix}}_{N \times N\text{ Matrix}} 
\end{equation}
\begin{equation}
  \text{and}\quad (\cos^2(\hat{\theta}))_{ij}(k) = \frac{1}{2}\underbrace{\begin{pmatrix}
1 & 0 & \sfrac{1}{2}& &  & & \dots& e^{-ik}/2& 0\\
0 & 1 & 0 & \sfrac{1}{2} & & & &\dots & e^{-ik}/2\\
\sfrac{1}{2} & 0 & 1 & 0 & \sfrac{1}{2}& & & & \\
& \sfrac{1}{2} & 0 & 1 & 0 & & & & \\
\vdots  &   &   & \ddots & & & & & \vdots \\
  &  &  &  & & &1 & 0&\sfrac{1}{2} \\
 e^{+ik}/2& &  & & & &0 & 1 &0\\
0 & e^{+ik}/2 & \dots & & & &\sfrac{1}{2} &0 & 1\\
\end{pmatrix}.}_{N\times N\text{ Matrix}} 
\end{equation}

\subsection{Appendix B: Symmetries}
Here, we discuss the symmetries of the model (following~\cite{PRL_charges}). As mentioned in the main text, the rotational phases, $\phi_{l}=e^{-\pi l(l+1)/N}$, give rise to reflection symmetry for all $N\in \mathbb{Z}$ for translationally-invariant potentials (\ie in our case for $l',l \gg 0$). For an odd $N$, the reflection center of the first unit cell is $n_c = (N+1)/2 \in \mathbb{N}$. Then, for a $0 \le j < n_c$ we have $\phi_{n_c + j} = \phi_{n_c - j}$ (for even $N$, the unit-cell is $2N$ and $n_c = N$ with $\phi_{n_c + j+1} = \phi_{n_c - j}$). If we limit ourselves to a finite Hilbert space with maximum $l_\mathrm{max}$, we can write the symmetry in the $l$-basis as
\begin{equation}
   \mathcal{P} = \underbrace{\begin{pmatrix} 
    & &\dots &0 & 1\\ 
    &  & & 1 & 0\\ 
  \vdots  &  & \iddots& & \vdots \\ 
  0 & 1& &&  \\ 
  1 & 0& \dots & \end{pmatrix}}_{(l_\mathrm{max}+1)\times(l_\mathrm{max}+1)\text{ Matrix}}
\end{equation}
with $\mathcal{P}^\dagger \mathcal{P} = \mathcal{P}^2 = \mathbb{I}$. It commutes with the time-translation operator, $[U, \mathcal{P}] = 0$. In terms of the effective Hamiltonian, $\hat{H}= \im \log \hat{U}$ [the Floquet period is absorbed in the Hamiltonian for simplicity], this symmetry reads in $k$-space $\mathcal{P}H_k = H_{-k}\mathcal{P}$ with
\begin{equation}
   \mathcal{P} = \underbrace{\begin{pmatrix} 
    & &\dots &0 & 1\\ 
    &  & & 1 & 0\\ 
  \vdots  &  & \iddots& & \vdots \\ 
  0 & 1& &&  \\ 
  1 & 0& \dots & \end{pmatrix}}_{N \times N \text{ Matrix}} \mathbb{P} \quad {\text{and the parity operator with}} \quad \mathbb{P}\phi_k = \phi_{-k}.
\end{equation}
For the one-kick system, shifting the position of the kick does not change the model, \ie
\begin{equation}
    U(\beta) = e^{-\im \hat{\mathbf{L}}^2\pi(1-\beta)/N }e^{\im V_1}e^{-\im \hat{\mathbf{L}}^2\pi\beta/N } \quad \sim \quad U = e^{-\im \hat{\mathbf{L}}^2\pi/N }e^{\im V_1}
\end{equation}
where the $\sim$ implies that the two operators have the same spectrum for any $0 \le \beta\le 1$. \\\\
In addition to that, we find a time-reflection symmetry present for all models with one kick in the Floquet operator (where we use that our laser potentials follow $V^*=V = V^T)$. More specifically, the second kick (for instance, $U_\text{2-kick} = e^{-\im \hat{\mathbf{L}}^2\pi/N}e^{\im V_1}e^{-\im \hat{\mathbf{L}}^2\pi/N}e^{\im V_2}$ with $V_1 \neq V_2$ ) would break the time-reflection invariance. The time-reflection invariance is proportional to unity times conjugation when the pulse is at the center, \ie~$\beta= \sfrac{1}{2}$. In that gauge, we find
\begin{equation}
   \langle l' |U^*(\sfrac{1}{2})U(\sfrac{1}{2})|l \rangle = \sum_{l''}\langle l'|e^{+\im \pi l'(l'+1)/2N}e^{-\im V}|l''\rangle e^{+\im \pi l''(l''+1)/2N}e^{-\im \pi l''(l''+1)/2N}\langle l''|e^{\im V}e^{-\im \pi l(l+1)/2N} |l\rangle = \delta_{l'l}
\end{equation}
which is equivalent to the time-reflection symmetry of the effective Hamiltonian $H(\beta)=H^*(\beta)$. To construct $\mathcal{T}$ for arbitrary gauges $\beta$, we only need to bring $U$ (or $H$, respectively) back to the $\beta= \sfrac{1}{2}$ gauge. This is accomplished by the gauge transformation $\mathcal{U}(\beta)$ with
\begin{equation}
   \mathcal{U}(\beta)=  e^{-\im \hat{\mathbf{L}}^2\pi(2\beta-1)/(2N)} \with \mathcal{U}^\dagger(\beta)U(\beta)\mathcal{U}(\beta) = U(\sfrac{1}{2}).
\end{equation}
The (anti-unitary) time-reflection operator then takes the form $\mathcal{T}(\beta) = \mathcal{U}(\beta) \, \mathcal{C}\, \mathcal{U}^\dagger(\beta)$ with conjugation $\mathcal{C}$. The time-translation operator then fulfills $\mathcal{T}(\beta)U(\beta)=U^\dagger(\beta)\mathcal{T}(\beta)$, while the Hamiltonian  $[H(\beta), \mathcal{T}(\beta)]=0$. In $k$-space, this leads to 
\begin{equation}
   \mathcal{T}(\beta) H_k(\beta) = H_{-k}^*(\beta)\mathcal{T}(\beta).
\end{equation}
Then, clearly any combination of time-reversal operators $\mathcal{U}_1,\mathcal{U}_M$ gives again rise to a new time-reversal operator by the symmetric combination
\begin{equation}
   U_\mathrm{sym}  =  U_1 \dots U_{M-1} U_M U_{M-1} \dots U_1
\end{equation}
and henceforth also the triple-kicked rotor with
\begin{equation}
U_{\mathrm{TKR}}(P) = {U}_{\mathrm{KR}}(P_1,P_2){U}_{\mathrm{KR}}(P_3,P_4){U}_{\mathrm{KR}}(P_1,P_2).
\end{equation}
Note that since this operator obeys both time-reversal and inversion symmetry, we can always gauge it to a real form. Let us define the eigenstates and eigenvalues of $\mathcal{P}$ by
\begin{equation}
    \mathcal{P}\mathcal{V}=\mathcal{V}\mathrm{diag}[\lambda],
\end{equation}
then, in the time-reversal symmetric frame ($\mathcal{T}=1$) the operator
\begin{equation}
    \mathcal{W} = \mathrm{diag}[\sqrt{\lambda}]\mathcal{V}^{\dagger}
\end{equation}
defines a transformation that gauges the Hamiltonian to be real, \ie
\begin{equation}
    H_k \mapsto \tilde{H}_k = \mathcal{W} H_k \mathcal{W}^\dagger \in \mathbb{R}^{N \times N}.
\end{equation}
Our analysis reveals that in the context of a time-reversal symmetric frame where inversion symmetry is preserved, a real gauge for the Hamiltonian can always be chosen~\cite{chen2021manipulation}. This holds true even when an auxiliary parameter is introduced, acting as a synthetic dimension that does not disrupt the existing symmetries.
Importantly, the system's $\mathcal{P}\mathcal{T}$ symmetry remains intact along an adiabatic path $P(\alpha)$, where $\alpha \in \mathbb{R}$, irrespective of any changes to these synthetic dimensions. This underscores the robustness of the $\mathcal{P}\mathcal{T}$ symmetry in our system.
However, it's crucial to clarify that the time-reversal symmetry in this context does not refer to physical time. Instead, it corresponds to an effective micro-motion within the Floquet cell. Consequently, the synthetic dimensions that might be varied over time are not subject to this particular time-reversal symmetry and hence they do not break it.
This nuanced understanding of time-reversal symmetry would be more accurately described as an 'anti-unitary symmetry akin to time-reversal'. Despite the complexity of its definition, one key property remains: this symmetry is conserved, which assures the system's classification within the Euler class and we confirm that eigenstates are real.
\subsection{$N=3$ case: Three-band case}
For $N=3$, the Fourier transform of the laser potential becomes 
\begin{equation}
\begin{aligned}
   V(k) &=  \frac{P_1}{2}\begin{pmatrix} 
    1 & e^{-ik}/2 & \sfrac{1}{2}\\
    e^{ik}/2 &1 & e^{-ik}/2 \\
    \sfrac{1}{2} & e^{ik}/2 &1
\end{pmatrix} 
+\frac{P_2}{2}\begin{pmatrix} 
    0 &1 &e^{-ik}\\
    1 &0 &1 \\
    e^{ik} & 1 &0
\end{pmatrix}  
\\ &= \frac{1}{2}\begin{pmatrix} 
    P_1 &P_1e^{-ik}/2+P_2 & P_1/2 + P_2e^{-ik}\\
    P_1 e^{ik}/2+P_2 &P_1 & P_1 e^{-ik}/2+P_2 \\
    P_1/2 +P_2 e^{ik} & P_1 e^{ik}/2+P_2 &P_1
\end{pmatrix}. 
\end{aligned}
\end{equation}
The reflection symmetry $1 \leftrightarrow 3, k \leftrightarrow -k$ is clearly visible here.

\section{Appendix C: Complete phase diagram}
In the main text, we presented only the nodal lines at $k=0$ and $k=\pi$ for the triple-kicked rotor for simplicity and better visibility. These lines differentiate different phases where the $\pi$-Zak (Berry) phases of the bands change. However, the triple-kicked quantum rotor indeed provides a fertile ground to realize various different nodal lines, phase transitions and braiding processes. This allows for a rich phase diagram with nodal lines that can be created at arbitrary quasi-momenta $k$ (and their symmetric counterparts at $-k$), as depicted in Fig.~\ref{fig:S2}. These nodal lines are created at symmetric momenta $k=0$ and $k=\pi$, branch out and form intricate nodal rings within the parameter space, which we emphasise can be controlled by tuning the external kicking parameters. In this context, we here demonstrate only the kicking strengths up to $P_{1,\mathrm{max}}=P_{2,\mathrm{max}} = 8$. 
\begin{figure}[htpb]
    \centering
    \includegraphics[width=1.0\textwidth]{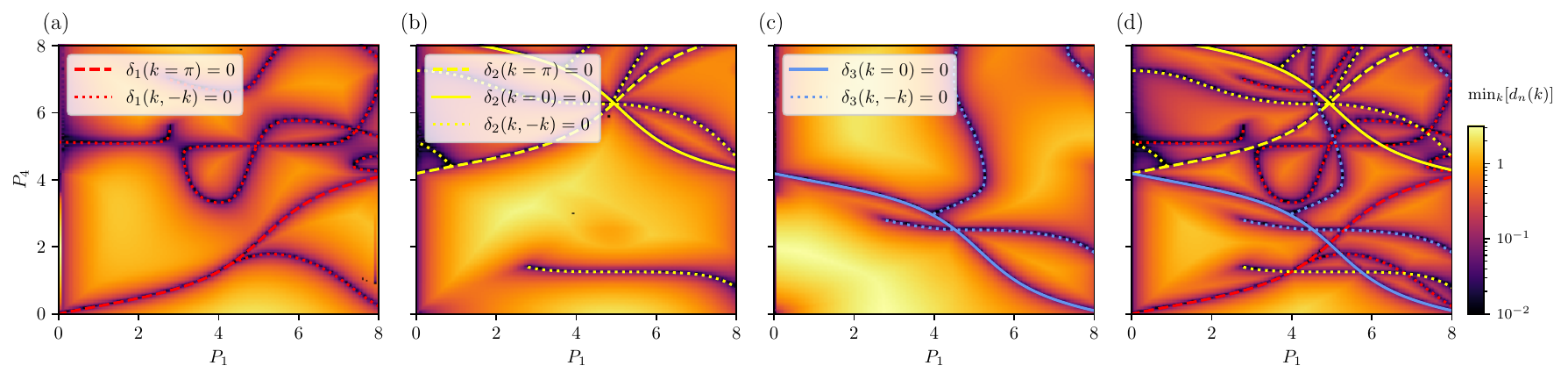}
    \caption{The nodal lines of the triple-kicked rotor when $P_2=P_3=0$. The solid and dashed lines represent the $k=0$ and $k=\pi$ nodal lines respectively, which were discussed in the main text. The dotted lines display the nodal lines for arbitrary momenta (at $(k,-k)$ pairs), thus providing a comprehensive view of the phase diagram. (a) Depicts the gap size (color-coded) of the first gap along with the nodal lines (in red) where the gap closes; (b) Illustrates the second gap (with gap closings in yellow); (c) Shows the third gap (with gap closings in blue); and (d) Presents the minimal gap size across all gaps, thereby providing a depiction of all the nodal lines within the system. We note that although the two-dimensional figures presented here might suggest otherwise, the nodal lines corresponding to different gaps do not touch each other; rather, they cross at different quasi-momenta $k$. As briefly discussed in the main text, this characteristic leads to a unique behavior of the non-Abelian topological charges associated with these nodal lines. When these lines (of different gaps) cross, their charges do not commute but instead undergo a sign change. Consequently, every time a dashed line (representing a specific gap) crosses a line of another gap (indicated by a different color), non-Abelian braiding occurs.}
    \label{fig:S2}
\end{figure}

\section{Appendix D: Calculation of the patch Euler class}
When evaluating the patch Euler class numerically, the calculation is slightly different from the usual evaluation of the Berry phase or Chern number, therefore we will explain it in the following with more detail (following~\cite{bouhon2019nonabelian}). First we calculate the linking matrix of the wavefunction, which represents the overlap between neighboring wavefunctions in the Brillouin zone grid. For a wavefunction $\psi_{n,\mathbf{k}}$, where $n$ is the band index and $\mathbf{k}$ is the wavevector, we define the linking matrices $U_x$ and $U_y$ as:
\begin{align}
U^x_{n}(k_x,k_y) &= \psi_{n,k_x,k_y} \cdot \psi_{n,k_x+\delta k_x,k_y} \label{eq:link1}\\
U^y_{n}(k_x,k_y) &= \psi_{n,k_x,k_y} \cdot \psi_{n,k_x,k_y+\delta k_y} \label{eq:link2}
\end{align}
where $\cdot$ denotes the inner product. To identify band nodes, we use the concept of Berry flux through an infinitesimal plaquette in the Brillouin zone. For each band $n$ and each plaquette centered at $(k_x, k_y)$, we calculate the Berry flux $\Phi_n(k_x, k_y)$ as:
\begin{equation}
\Phi_n(k_x, k_y) = \text{sign}\left(U^x_n(k_x,k_y)  U^x_n(k_x,k_y+\delta k_y)  U^y_n(k_x,k_y)  U^y_n(k_x+\delta k_x,k_y)\right).
\end{equation}
A band node is identified when $\Phi_n(k_x, k_y) = -1$, showing a $\pi-$Berry phase around the plaquette. This identification stems from the fact that each band node generates a Dirac string, which leads to a $\pi-$Berry phase when a contour encircles the band node. In the framework of topological band theory, Dirac strings are conceptualized as singular lines in the Brillouin zone, connecting pairs of band nodes (referred to as Dirac points in this context). As emphasized in the main text, it is important to note that these strings are not physical observables, but rather gauge-dependent constructs. They emerge from the necessity to establish a smooth gauge for the wavefunctions across the entire Brillouin zone. A contour that does not encompass a band node inevitably intersects Dirac strings an even number of times, resulting in no net Berry flux. This property ensures that the Berry flux is a robust indicator of the presence of band nodes, independent of the specific gauge choice. Henceforth, we can choose the Dirac strings that connect the band nodes in a for us convenient way and fix the gauges of the wavefunction accordingly, such that it is smooth almost everywhere, except when crossing a Dirac string, where the sign is flipped. Then, we can evaluate the Euler curvature which characterizes the local topology of the band structure.
First, we complexify the two wavefunctiosn $\phi(\mathbf{k}) = \frac{1}{\sqrt{2}}(\psi_1(\mathbf{k}) + i\psi_2(\mathbf{k}))$ in the Euler subspace, which gives again rise to linking matrices $\mathcal{U}^{x,y}(k_x,k_y)$, analogous to~\eqref{eq:link1} and~\eqref{eq:link2}. For each plaquette in the Brillouin zone that does not contain a band node, we calculate the Euler form $\mathrm{Eu}(k_y,k_y)$ as follows
\begin{equation}
    \mathrm{Eu}(k_x,k_y)\mathrm{d}k_x\mathrm{d}k_y = -\mathrm{Arg}\left[
    \mathcal{U}^y(k_x,k_y)
    \mathcal{U}^x(k_x,k_y+\delta k_y)
    \mathcal{U}^{y*}(k_x+\delta k_x,k_y)
    \mathcal{U}^{x*}(k_x,k_y)
    \right].
\end{equation}
For plaquettes that contain band nodes, we need to apply an appropriate gauge transformation (\ie multiplication by $(-1)$) that ensure that the linking matrix is continuous in that specific direction and analytically continue the curvature by 
\begin{equation}
\begin{split}
\tilde{\mathrm{Eu}}(k_x, k_y) = & \frac{1}{6} \left[\mathrm{Eu}(k_x+\delta k_x, k_y) + \mathrm{Eu}(k_x-\delta k_x, k_y) + \right. \\
                & \left. \mathrm{Eu}(k_x, k_y+\delta k_y) + \mathrm{Eu}(k_x, k_y-\delta k_y)\right] + \\
                & \frac{1}{12} \left[\mathrm{Eu}(k_x+\delta k_x, k_y+\delta k_y) + \mathrm{Eu}(k_x-\delta k_x, k_y+\delta k_y) + \right. \\
                & \left. \mathrm{Eu}(k_x+\delta k_x, k_y-\delta k_y) + \mathrm{Eu}(k_x-\delta k_x, k_y-\delta k_y)\right]
\end{split}
\end{equation}
This correction term is a weighted average of the Euler curvature values at the neighboring points, with more weight given to the nearest neighbors (factor of 1/6) compared to the diagonal neighbors (factor of 1/12). This approach ensures a smooth interpolation of the Euler curvature at band nodes. As a next step we need to calculate the contour integral of the non-Abelian connection $\mathcal{A}=\langle \psi_1|\nabla_\vec{k} \psi_2 \rangle$ around the chosen boundary of the patch. The integral is discretized and computed along the four edges of the patch similar to the Berry phase calculation for the Zak phases. Finally, the calculation of the Euler invariant proceeds with the subtraction of the two terms with 
\begin{equation}
   \chi_n = \frac{1}{2\pi} \left ( \int_\mathrm{D} \mathrm{Eu}(k_x,k_y) \mathrm{d}k_x \wedge \mathrm{d}k_y- \int_{\partial \mathrm{D}}\mathcal{A} \right ) \in \mathbb{Z}.
\end{equation}

\end{widetext}

\end{document}